\documentclass[a4paper,fleqn,usenatbib, letters]{mnras}

\usepackage{newtxtext,newtxmath}

\usepackage[T1]{fontenc}
\usepackage{ae,aecompl}

\usepackage{graphicx}	
\usepackage{amsmath}	
\usepackage{amssymb}	

\title[Detection of helicoidal motion in the optical jet of PKS\,0521$-$365]{Detection of helicoidal motion in the optical jet of PKS\,0521$-$365}

\author[E. F. Jim\'enez-Andrade et al.]{
E. F. Jim\'enez-Andrade,$^{1,2,3}$\thanks{E-mail: ericja@astro.uni-bonn.de}
 V. Chavushyan,$^{3}$ 
 J. Le\'on-Tavares,$^{4}$
\newauthor  V. M. Pati\~no-\'Alvarez,$^{5}$ A. Olgu\'in-Iglesias,$^{3}$ J. Kotilainen,$^{6,7}$ R. Falomo$^{8}$  \newauthor and   T. Hyv\"onen$^{9}$
\\
$^{1}$Argelander-Institut f\"ur  Astronomie, Universit\"at Bonn, Auf dem H\"ugel 71, D-53121 Bonn, Germany\\
$^{2}$International Max Planck Research School of Astronomy and Astrophysics at the Universities of Bonn and Cologne \\
$^{3}$Instituto Nacional de Astrof\'isica \'Optica y Electr\'onica (INAOE), Apartado Postal 51 y 216, 72000 Puebla, M\'exico\\
$^{4}$Centre for Remote Sensing and Earth Observation Processes (TAP). Flemish Institute for Technological Research (VITO), \\ Boeretang  282, 2400 Mol, Belgium\\
$^{5}$Max-Planck Institut f\"ur Radioastronomy, Auf dem H\"ugel 69, 53121 Bonn, Germany\\
$^{6}$Finnish Centre for Astronomy with ESO (FINCA), University of Turku, V\"ais\"al\"antie 20, 21500 Kaarina, Finland\\
$^{7}$Tuorla Observatory, Department of Physics and Astronomy, University of Turku, V\"ais\"al\"antie 20, 21500 Kaarina, Finland\\
$^{8}$Osservatorio Astronomico di Padova, INAF, vicolo dell'Osservatorio 5, 35122 Padova, Italy\\
$^{9}$Faculty of Natural Sciences, Tampere University of Technology, P.O. Box 589 FI-33101 Tampere Finland
}

\date{Accepted XXX. Received YYY; in original form ZZZ}

\pubyear{2017}

\begin{document}
\label{firstpage}
\pagerange{\pageref{firstpage}--\pageref{lastpage}}
\maketitle

\begin{abstract}
 The jet activity of Active Galactic Nuclei (AGN), and its interaction with the interstellar medium (ISM), may play a pivotal role in the processes which regulate the growth and star formation of its host galaxy. Observational evidence which pinpoints the conditions of such interaction is paramount to unveil the physical processes involved. We report on the  discovery of  extended emission  line regions exhibiting an S-shaped morphology along the optical jet of the radio-loud AGN PKS\,0521$-$365 ($z=0.055$), by using  long-slit spectroscopic observations obtained with FORS2 on VLT. The velocity pattern derived from the  [O\,II] $\lambda 3727$\,\AA, H$\beta \ \lambda 4861$\,\AA \ and [O\,III] $\lambda\lambda4959,5007$\,\AA\, emission lines is well-fitted by a sinusoidal function of the form: $v(r)=\alpha r^{1/2}sin(\beta r^{1/2}+\gamma)$, suggesting helicoidal motions along the jet up to distances of 20\,kpc.  We estimate a lower limit for the mass of the outflowing ionized gas along the jet of $\sim$10$^4\,M_\odot$. Helical magnetic fields and jet precession have been proposed to explain helicoidal paths along the jet at pc scales; nevertheless, it is not clear yet whether these hypotheses may hold at kpc scales. 
\end{abstract}




\section{Introduction}

\begin{figure*}
	\centering
	\includegraphics[width=0.7\linewidth]{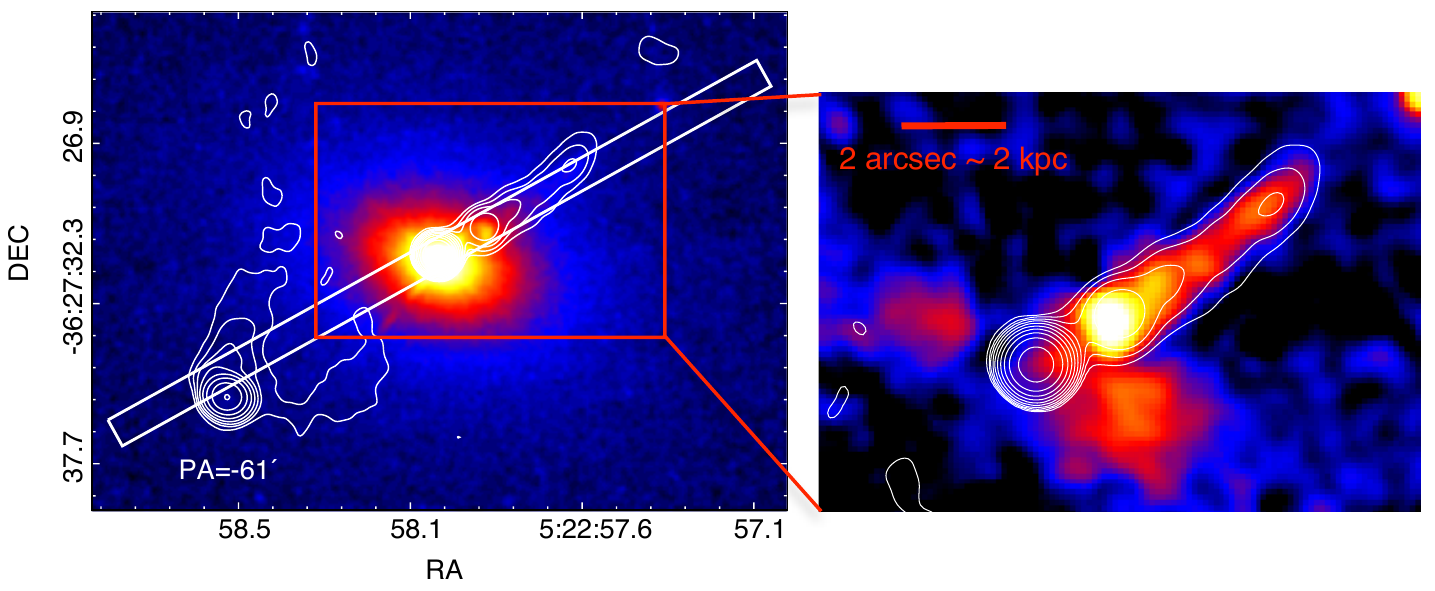}	
	\caption{{\it Left panel:} Optical HST image (WFPC2/F702W) of PKS\,0521$-$365 which  features the host galaxy emission and the prominent jet. The contours represent the VLA radio map at 15 GHz \citep[contour levels: -1, 1, 2, 4, 8, 16, 32, 64, 128, 256, 512 mJy beam$^{-1}$; ][]{Falomo2009}. Solid white rectangle shows the position of the slit (PA=$-$61.0$^\circ$). {\it Right panel:}  Residual image obtained after modeling and subtracting  the host galaxy and nucleus contribution with {\sc Galfit}. The diffuse emission along the major axis of the host galaxy  might be closely related with the  structure perpendicular to the radio jet  detected with ALMA \citep{Leon2016}.  The contour levels correspond to the VLA radio map at 15\,GHz described above. }
	\label{fig1}
\end{figure*}

The energy released by AGN is thought to significantly impact the evolution of its host galaxy \citep[e.g.][]{Fabian2012, kormendy13, king15}. Observational evidence of winds driving gas out of the nuclear regions \citep[e.g.][]{Holt08, Harrison2012, Combes13,  Morganti2013, Morganti13b, Dasyra2015, Morganti15, Collet16, Querejeta16} suggests that outflows might be the main mechanism that could efficiently transfer energy from scales close to the black hole (pc) to host galaxy scales (kpc). These outflows -- which arise as a by-product of accretion onto a black hole -- are usually associated with either an accretion disk or radio jets \citep[e.g.][]{Croton2006,  Krongold2007}; nevertheless, the physical processes which regulate the interplay between the radio jet activity and the multi-phase gas remain unclear. 
On this regard, spatially resolving the interaction between an AGN and its host galaxy will provide key constraints on the physics and ubiquity of AGN feedback. For instance, long-slit spectroscopic studies have proved to be well suited to resolve jet-cloud interactions in nearby radio galaxies, suggesting strong interactions between the radio-emitting plasma and the ISM  \citep[e.g.][]{Clark97, Armus98, Clark98, VillarMartin1999, Emonts2005, Inskip2008, Rosario2010}.  On the other hand, Integral Field Spectroscopy (IFS) is particularly useful to  disentangle the kinematic components and ionization state of complex extended emission line regions which are not only limited to the radio jet axis \citep[e.g.][]{Solorzano2003,Inskip2008,Santoro2015}.

PKS\,0521$-$365 is one of the most studied radio-loud AGN in the southern sky. Yet, there is no robust observational evidence on the effect of the  AGN activity on its host galaxy \citep[e.g.][]{Hyvonen2007}. Therefore it becomes one of the most accessible targets (z = 0.05548) for spatially resolving  the trace that powerful jets leave on its host galaxy. Classified as a Flat Spectrum Radio Quasar (FSRQ) \citep{Scarpa1995}, it shows a large-scale optical/near-IR/sub-mm jet well aligned with the kpc radio jet \citep[][see Fig.~\ref{fig1}]{Scarpa1999, Falomo2009, Leon2016}.  Recently, a diffuse and extended structure perpendicular to the radio jet was detected in  bands 3 and 6 with ALMA,  which may be related with the relic of a previous jet or thermal (dust) emission associated with a central star-forming region \citep{Leon2016}.   In this letter  we report the discovery of extended emission  line regions  exhibiting an S-shaped morphology which suggests helicoidal motions along the jet of PKS\,0521$-$365 at kpc scales, providing new evidence on the way that AGN jets interact with the ISM. We report in Section 2 the details of the observations, followed by the results and analysis in Section 3.  A discussion is given in Section 4.  A cosmology with H$_0$=70\,km\,s$^{-1}$ Mpc$^{-1}$, $\Omega_m$=0.30, and  $\Omega_\Lambda$=0.70 is assumed, corresponding to a luminosity distance for PKS\,0521$-$365 of 247.6\,Mpc and a scale of 1.078\,kpc arcsec$^{-1}$.

\section{Observations}

We  secured  long-slit spectra along the direction of the optical jet of PKS\,0521$-$365 (PA=$-61.0^\circ$, see Fig.~\ref{fig1}) with the Very Large Telescope (VLT), using  the FOcal Reducer/ low dispersion Spectrograph 2 \citep[FORS2][]{Appenzeller1998} and the GRIS 600B+22 (wavelength range 3300-6210  \AA, dispersion 50  \AA/mm).  Three consecutive spectra of 850\,s integration time each were obtained during December 12, 2008 under good atmospheric conditions  (seeing$\sim$0.7; air mass$\sim$1.1). The data reduction was performed using the standard procedures with {\sc Iraf}.    In the first stage, bias subtraction, flat fielding and removal of bad pixels were applied. Then, wavelength calibration, background subtraction  were performed before combining the three spectra into a single spectrum. Flux calibration was performed after extracting 1D spectra along the spatial axis by using the standard star LT2415B. This result in a long-slit spectrum which encompass emission from the central engine and the optical jet  with adequate spectral resolution ($\text{\sc FWHM}_{\text{sky-lines}}$=4.5\,\AA) and high S/N ($\sim$100).

To pinpoint the spatial region covered by our long-slit spectroscopic data, we also use the HST image of PKS\,0521$-$365 using WFPC2 in the R(F702W) filter \citep{Scarpa1999}.  The optical image was modeled using the galaxy fitting algorithm {\sc Galfit}. We use the point spread function (PSF) model, obtained with the HST PSF modeling tool Tiny Tim,  to represent the nuclear region of the galaxy. Similarly, we used a S\'ersic profile convolved with the PSF to represent the host galaxy. According to our analysis, the host galaxy of PKS\,0521$-$365 is (as expected) a giant elliptical with a S\'ersic index $n$=3.96$\pm$0.41, an effective radius $R_e$= (4.74$\pm$0.55)\,kpc, an ellipticity {\it E}=0.23$\pm$0.10 and a magnitude {\it m}=18.13$\pm$0.52 ({\it M}=$-$18.88$\pm$0.52), well in accordance with previous analysis on PKS\,0521$-$365  \citep{Urry2000} and typical values of blazars hosts \citep{OlguinIglesias2016}.  In the right panel of Fig.~\ref{fig1}, we show the optical HST image when we subtract the modeled nucleus and host galaxy,  revealing an optical jet which displays knotty morphologies and  reassembles the structure of the radio jet.

\section{Results}

\begin{figure}
	\centering		
	\includegraphics[width=0.85\linewidth]{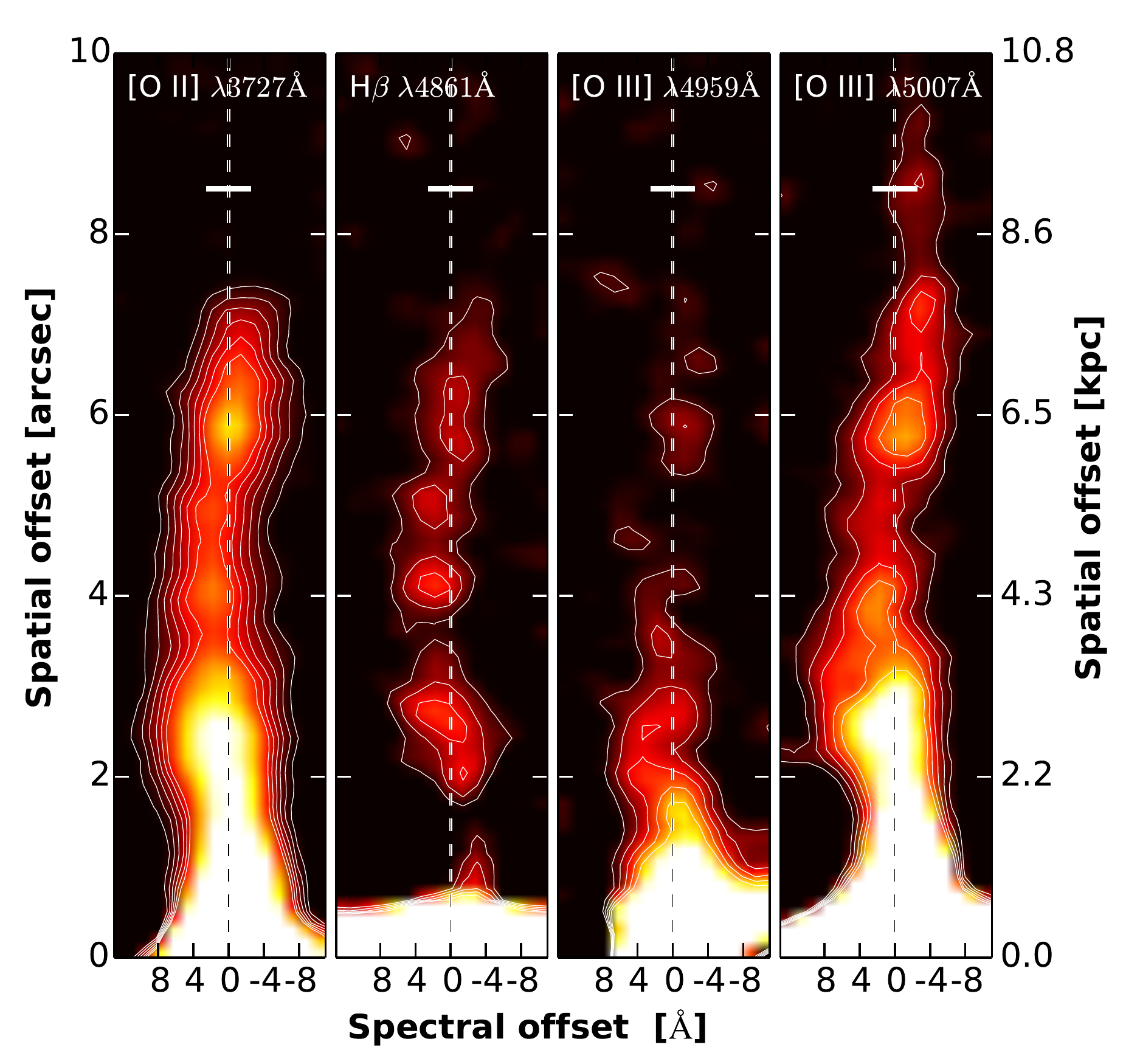}
	\caption{ 2-D images showing extended emission lines revealed by  long-slit spectra obtained with the slit oriented along the optical jet (PA=$-61.0^\circ$) --  continuum emission has been  subtracted. The extended emission lines [O\,II] $\lambda 3727$\,\AA, H$\beta \ \lambda 4861$\,\AA \ and [O\,III] $\lambda\lambda4959,5007$\,\AA,  feature an S-shaped morphology (contour levels: 3, 6, 9, 12, 15, 18$\sigma$). Our spectral resolution, given by the FHWM from the sky-lines of 4.5\,\AA, is shown in all the panels as a white horizontal line.}
	\label{fig2}
\end{figure}

\begin{figure}
	\centering
	\includegraphics[width=0.85\linewidth]{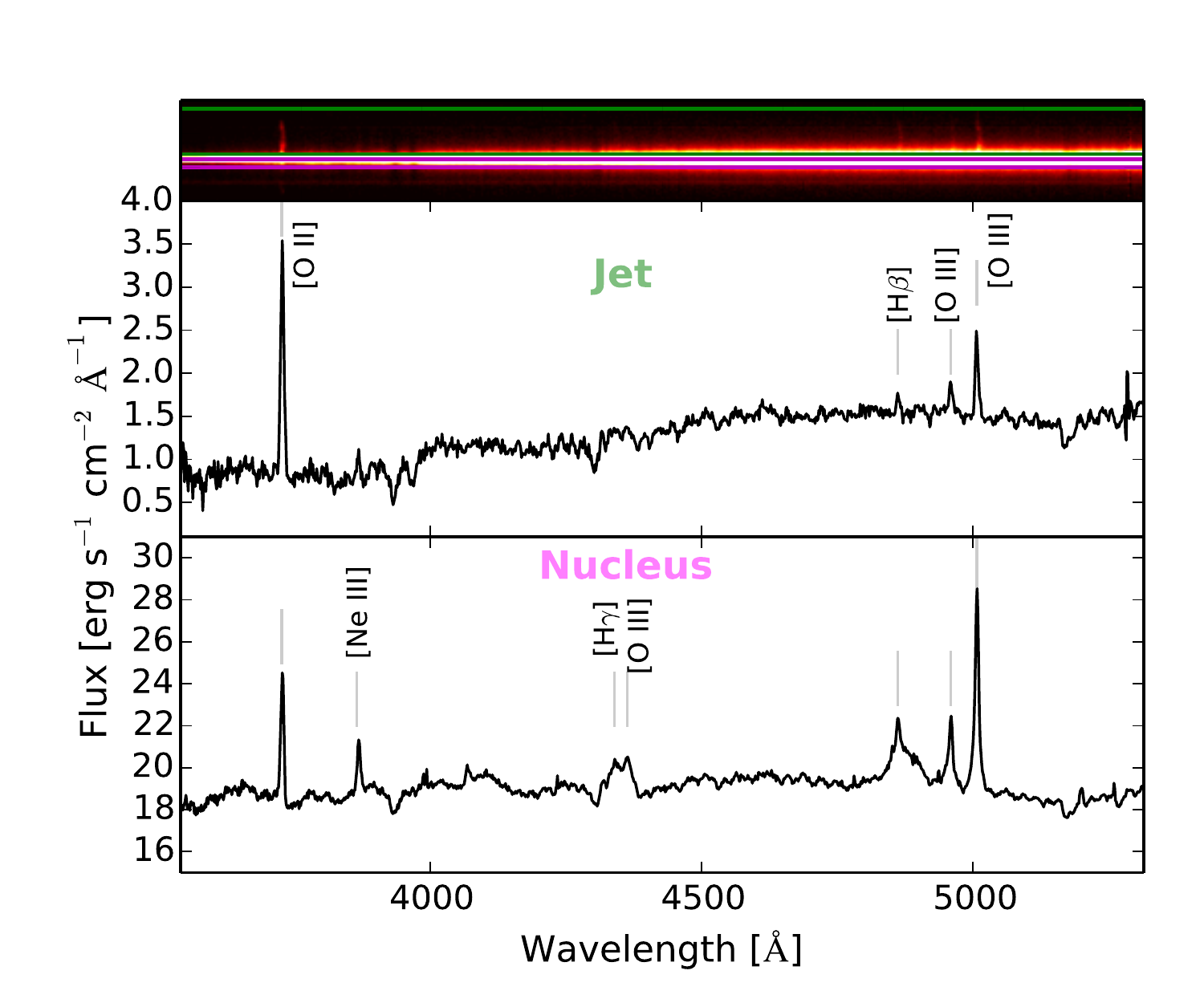}	
	\caption{ {\it (Upper panel)} 2D spectrum of PKS\,0521$-$365, green and magenta lines show the regions  where the emission from the optical jet and the central engine was integrated.  {\it (Middle panel)} Spectrum from the optical jet which exhibits narrow emission lines. {\it (Lower panel)} Spectrum from the inner kpc in  PKS\,0521$-$365 which shows strong emission lines from the central engine  and shallow stellar absorption lines.}
	\label{fig3}
\end{figure}

Long-slit spectroscopy has  revealed a  large number of extended emission line regions aligned with the radio jet axis of radio-loud AGN. The spatial extent, ionization state and velocity fields of these regions have been examined in some detail \citep[e.g.][]{Best1997, Scarpa1999, VillarMartin1999, Emonts2005, Rosario2010b, Rosario2010, Liuzzo2011}. 
In this work, we report on the finding of extended  emission line regions along the optical jet of PKS\,0521$-$365.  In the 2D spectrum (see Fig.~\ref{fig1},~\ref{fig2}), the warped and knotty emission lines [O\,II] $\lambda 3727$\,\AA, H$\beta \ \lambda 4861$\,\AA \ and [O\,III] $\lambda\lambda4959,5007$\,\AA\, spread along the spatial axis towards the direction of the optical jet, which suggests ongoing jet-cloud interactions. In fact, this emission  corresponds to the emitting knots traveling along the jet revealed by the HST imaging (see Fig.~\ref{fig1}).
We do not detect extended emission in the direction of the counter-jet in our long-slit spectroscopic nor optical imaging data (see Fig.~\ref{fig1},~\ref{fig3}); which might be due to the fact that relativistic beaming enhance  the approaching jet flux and dim the receding one \citep{Leon2016}.

The optical 1D spectra from the  optical jet emission and the central engine were obtained by co-adding  emission along the spatial axis in the 2D spectrum (see Fig.~\ref{fig3}). In the first case, we integrate emission spreading up to $\sim$10 arcsec from the nucleus and neglect that from  the inner $\sim$1.5 arcsec which is contaminated by broad line emission.  In physical units, this  corresponds to an area of $\sim$8.5\,kpc\,$\times$\,1\,kpc, without correcting for the jet orientation. Similarly, to get the spectrum from the central engine we integrate emission from the inner kpc.  As expected, the spectrum of the  optical jet emission exhibits strong  [O\,II] $\lambda 3727$\,\AA\, and [O\,III] $\lambda5007$\,\AA\,  narrow emission lines as well as stellar absorption features.  On the other hand, the spectrum from the nucleus is dominated by the central engine with strong and broad  H$\beta \ \lambda 4861$\,\AA\, line emission along with narrow emission lines.

The spectral coverage of the spectra (3500$-$6000\AA\,) does not allow to detect the [O\,I] $\lambda 6300$\,\AA, H$\alpha \ \lambda 6563$\,\AA\,  and  [S\,II] $\lambda\lambda6717,6731$\,\AA\,  emission lines; which are essentials to probe the physical conditions (density, temperature) and disentangle the ionization mechanism of the emitting gas by using nebular emission line diagnostic diagrams (e.g. BPT diagrams). Thus, probing the ionization state of the gas is beyond the scope of this work. Nevertheless, we profit from the intermediate spectral resolution and high S/N of our spectra to examine the  radial velocity patterns of the diffuse and warped emission lines spreading along the optical jet of PKS\,0521$-$365 (see Fig.~\ref{fig2},~\ref{fig3}). 

\subsection{Velocity profile}

To explore the kinematics  of the gas clouds along the optical jet we  extract their spatial profile  from the 2D spectrum. 
We first remove the ``contamination'' by adjacent continuum emission in the 2D spectrum (which spread over a few arcsec along the spatial axis) with the task \emph{continuum} in {\sc Iraf}. 
We integrate the emission (detected above 3-$\sigma$ in the 2-D spectrum) along the spatial axis in bins of 5 pixels -- our spatial resolution is $\sim$4.5 pixels, where 1\,pixel\,=\,0.126\,arcsec.  A single Gaussian function is fitted to each line in order to obtain the  amplitude, $\sigma$ and central wavelength of the line profile; from the latter parameter, we estimate the velocity offset with respect to the systemic velocity of the host galaxy. Given that stellar features in the spectrum from the nucleus are shallow, we use narrow emission lines ([O\,II] $\lambda 3727$\,\AA, [O\,III] $\lambda\lambda4959,5007$\,\AA) associated with the central engine to derive the systemic velocity.  We do not use any constraint on the separation nor line ratio of the [O\,III] doublet in order to obtain independent measurements. It should be noted that the spectral resolution, given by the FWHM from the sky-lines (FWHM$_{\text{sky-lines}}$=4.5\,\AA), suffices to resolve the emission spreading within a spectral range of $\sim$15\,\AA\, (in the case of [O\,II] $\lambda 3727$\,\AA\, and  [O\,III] $\lambda5007$\,\AA).

\begin{figure*}
	\centering		
	\includegraphics[width=0.9\linewidth]{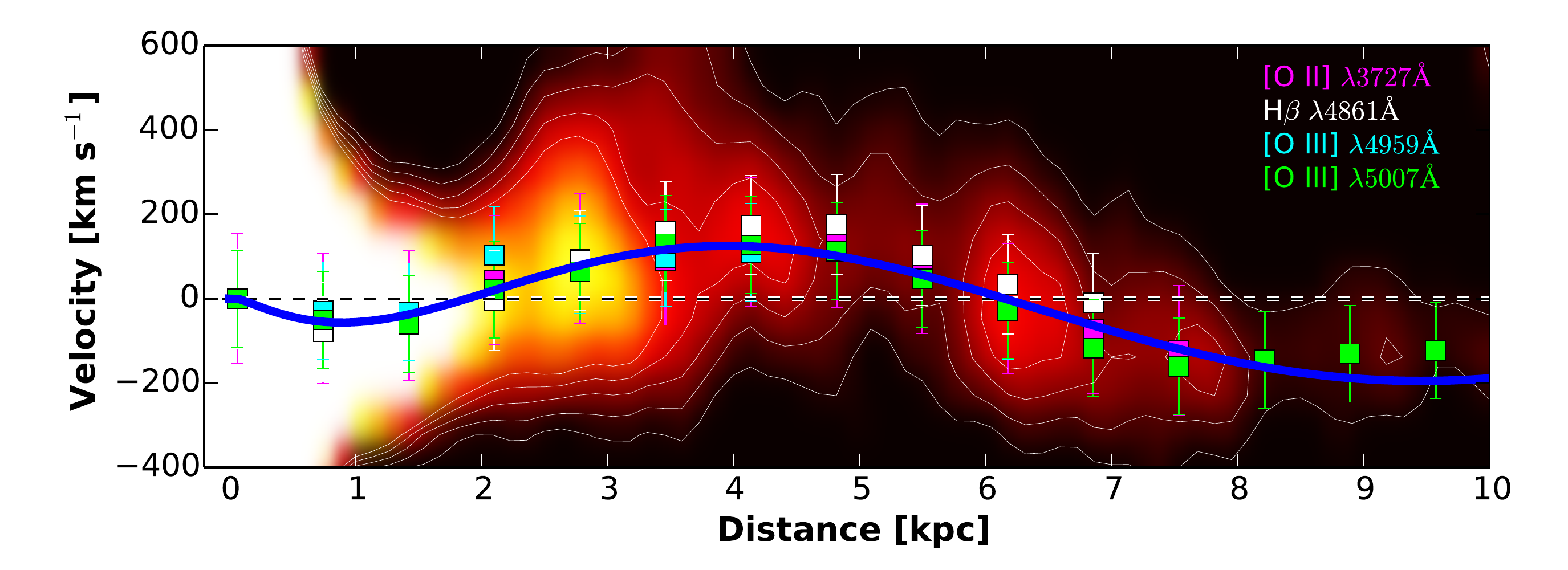}
	\caption{Velocity profile along the optical jet (PA=$-61.0^\circ$) of PKS\,0521$-$365 derived from the [O\,II] $\lambda 3727$\,\AA, H$\beta \ \lambda 4861$\,\AA \ and [O\,III] $\lambda\lambda4959,5007$\,\AA\, emission lines. The background image shows smoothed emission of the [O III] $\lambda 5007$\,\AA \, emission line in the 2D spectrum, while the squares show the derived profiles after binning the emission along the spatial axis. The solid blue line represents the sinusoidal model to fit the data. The length of the error bars is given by the spectral resolution, i.e. $\sigma_s=\text{FWHM}_{\text{sky-lines}}/(2\sqrt{2\ln(2)})\sim 1.8$\AA.  
		The wavelength axis has been labelled  in velocity units with respect to the systemic velocity of the galaxy; while the spatial axis has been labelled  in physical distance units (kpc) according  to the given plate scale (0.126 arcsec/pixel) and the assumed cosmological parameters.} \vspace{0.5cm}
		\label{fig4}
\end{figure*}

The velocity profile along the optical jet derived from the four lines is shown in  Fig.~\ref{fig4}. Although the velocity swings are evident in the 2D spectrum, we perform a chi-square goodness of fit test to explore whether our data points can be described by a constant function, $v(r)=v_c$;  where  $v_c$ is the mean velocity in km\,s$^{-1}$ along the spatial axis. We derive $\chi^2=28$ -- with $44$ degrees of freedom -- which yield a \emph{p-value} of 0.035. Consequently, since the \emph{p-value} is smaller than the significance level (0.05) we can reject the null hypothesis, meaning that the data is not consistent with  a constant function.   On the other hand, it should be noted that the velocity  profiles remarkably resemble an S-shape suggesting a sinusoidal behavior. Thus, for fitting the data better than a linear model we propose a  sinusoidal function (defined by three coefficients) to fit the velocity profile:

\begin{equation}
v(r)=\alpha r^{1/2}sin(\beta r^{1/2}+\gamma)
\end{equation}

where $v(r)$ is the velocity in km\,s$^{-1}$, $r$ is the distance in kpc and $\alpha,\beta$ and $\gamma$ are constants to be determined. We use a non-linear least-squares (Levenberg-Marquardt) algorithm to  find  the best-fitting values for these constants: $\alpha=(64\pm4)\,\text{km}\,\text{s}^{-1}\,\text{kpc}^{-1/2}
,\ \beta=(-2.8\pm0.12)\,\text{kpc}^{-1/2}$ and $\gamma=-11.8\pm0.3$. This function describes a sinusoidal movement of ionized matter along the jet; where both, amplitude and period, increase with the distance. The farthest detected emission lies at 10 kpc -- without correcting by the jet orientation -- and the projected velocity  reaches a maximum of  200 km\,s$^{-1}$. In fact, under the conservative assumption of having a viewing angle of 30$^{\circ}$ \citep{Pian1996, Giroletti2004} the optical emission along the jet would extend up to 20 kpc.

In order to discern if the observed velocity shifts in the emission lines are consistent with the proposed model, we apply a Kolmogorov-Smirnov (K-S) test \citep{Press1986}.  We simulated a sample of distance values by Monte Carlo simulations, then velocity shift values were obtained from the sinusoidal function, for each simulated value. By comparing the observations to the generated sample drawn from a distribution based on the sinusoidal model, we obtained a K-S statistic of 0.15  and a significance level of the K-S statistic of 0.61. Such high significance level points towards the null hypothesis being correct. From the K-S analysis, we conclude that both samples, observed and simulated, are drawn from the same parent distribution; which strengthens the argument of the sinusoidal motion along the jet.

\subsection{Mass outflow along the jet}

If the gas in a line-emitting region is primarily photoionized,  the mass of the gas  can be estimated from the  H$\beta$ luminosity as follows \citep{Osterbrock1989},

\begin{center}
\begin{equation}
M_{\text{gas}}=m_p\frac{L(\text{H}\beta)}{n_e\alpha^{eff}_{\text{H}\beta}h\nu_{\text{H}\beta}}
\end{equation}
\end{center}

where $n_e$ is the electron density in cm$^{-3}$, $m_p$ is the mass of a proton  in kg, $L$(H$\beta)$ is the H$\beta$ luminosity in \ erg\ s$^{-1}$, $\alpha^{eff}_{\text{H}\beta}$ is the effective recombination coefficient for H$\beta$ in cm$^{3}$ s$^{-1}$ and $h\nu_{\text{H}\beta}$ is the energy of an H$\beta$ photon in erg. We are assuming $T=10000$\,K, since this is a typical temperature for a photoionized line emitting region \citep{Osterbrock1989}.

We apply the {\sc Starlight} code \citep{Cidfernandes2005}  to the spectrum from the  optical jet emission to subtract the host galaxy contamination and AGN non-thermal continuum.  Before running {\sc Starlight} the spectrum was corrected for Galactic extinction assuming the E$_{\text{B-V}}$ values computed by \cite{Schlegel1998}.  We estimate $F_{\text{H}\beta}\sim5 \times 10^{-16}$\,erg\,s$^{-1}$\,cm$^{-2}$; this value should be considered as a lower limit since the width of the slit itself (1\,arcsec) does not allow to recover the entire extended emission along the jet.  We consider typical densities of extended emission-line regions aligned with the radio axis which have been reported in the literature \citep[e.g.,][]{Emonts2005, Nesvadba2006, Nesvadba2008, Rosario2010}. In general, derived values for the density in extended emission-line regions -- where the jet is strongly interacting with the ISM -- range from  200 to 1000\,cm$^{-3}$, which yields a mass of  $10^{5.1}\text{M}_\odot$ and $10^{4.4} \text{M}_\odot$ respectively.  This result is in agreement with previously reported masses of jet-induced outflows of ionized gas, which are of the order of $10^{4-5} \text{M}_\odot$ \citep{Emonts2005, Rosario2010b, Rosario2010}. Naturally, these values are considered as lower limits for the total outflowing gas mass, as neutral and molecular gas might be present as well.

\section{Discussion}

We found ordered  emitting gas motions along the  jet of the active galaxy  PKS\,0521$-$365 with a mass of at least $10^4\,M_\odot$.
Evidence of bright optical knots tightly aligned along the jet in radio galaxies has been already reported in the past; for instance, in 3C\,266, 3C\,324, 3C\,368, 3C\,371, PKS\,2201$+$044 and PKS\,2250$-$41  \citep{Best1997, Scarpa1999, VillarMartin1999, Liuzzo2011}. Such alignments suggest that strong interactions are taking place between the jet and the line emitting gas which might derive into jet-triggered star formation \citep{Tremblay2015, Donahue2015}. 
In this work, we report the finding of narrow-line emitting gas oriented along the jet of PKS 0521$-$365 and  provide insights about the kinematic of these regions. We found that the gas radial velocity patterns can be well described by a sinusoidal function, giving the first  spectroscopic evidence of helicoidal motions along the jet on kpc scales. 

Very Long Baseline Interferometry (VLBI) studies have revealed that helical structures are common in extragalactic jets in pc scales \citep[e.g.][]{Lister2003}.  They are usually associated with helical magnetic fields  which are linked to the rotation of the central black hole and its accretion disk together with the jet outflow \citep{Steffen1995, Keppens2008}. On the other hand, helical structures may be a consequence of  jet precession caused  by a supermassive binary black hole system  (SBBH) or  the accretion disk;  the gas accretion -- possibly driven by minor mergers -- is likely to occur at random angles \citep{Roos1993, Ostorero2004, Lu2005, Aalto2016}. Hence, the S-shaped jet morphologies may  reflect the fact that their black hole spin axis is still precessing and has not had sufficient time to align with the accretion disk. In particular,  the presence of a  SBBH or recent merging activity in PKS\,0521$-$365 remains an open question.
The later models have proved  to be well suited to the observations in pc scale jets, nevertheless, what remains puzzling is at  what extension these models can predict  a helical path. Further theoretical and observational studies are needed to reconcile these approaches with the extent of the kpc-scale jet of PKS 0521$-$365 which shows signs of helical structures.

PKS\,0521$-$365 is a multifaceted object which is undergoing a high-energy episode. It represents a unique opportunity to further inspect in detail the kinetic influence and ability of radio jets to drive gas outflows and interact with the ISM of its host galaxy; in particular, to understand how radio jets can transfer energy and redistribute mass up to galactic scales, and whether they can drive star formation in the timescales that they are acting upon the gas.

\section*{Acknowledgements}

EFJA acknowledge support  from the CONACyT (M\'exico) Master's programs  and from the Collaborative Research Council 956, subproject A1, funded by the Deutsche Forschungsgemeinschaft (DFG, Germany). This work was partially supported by CONACyT research grant 151494. The observations used in this study were carried out at the European Southern Observatory (Paranal, Chile) with FORS2 on VLT (program  82.B-0720(A), PI: T. Hyv\"onen).





\bibliographystyle{mnras}
\bibliography{Jimenez-Andrade+.bib} 







\bsp	
\label{lastpage}
\end{document}